\documentclass[adp,fleqn]{w-art}
\usepackage{times}
\usepackage[]{graphicx}

\usepackage{amssymb}
\usepackage{amsmath,bm}

\def\a{\alpha}
\def\b{\beta}
\def\g{\gamma}
\def\d{\delta}

\def\vt{\vartheta}

\begin{document}

\DOIsuffix{theDOIsuffix}
\Volume{XX}
\Issue{x}
\Copyrightissue{x}
\Month{01}
\Year{2008}
\pagespan{1}{}
\Receiveddate{25 July 2008}
\Dateposted{{\it file MinkowskiEdyn8.tex}}

\title[Maxwell's equations in Minkowski's world]{Maxwell's
    equations in Minkowski's world:\\their premetric
    generalization and\\the electromagnetic
    energy-momentum tensor}

\author[Friedrich W.~Hehl]{Friedrich
  W.~Hehl\footnote{E-mail:~\textsf{hehl@thp.uni-koeln.de}} }

\address[]{Institute for Theoretical Physics, University of
  Cologne, 50923 K\"oln, Germany   \\{\it and} \\
  Department of Physics and Astronomy, University of
  Missouri-Columbia, Columbia, MO 65211, USA} 

\begin{abstract}
  In December 1907, Minkowski expressed the Maxwell equations in the
  very beautiful and compact 4-dimensional form: lor $f=-s,\;$ lor
  $F^*=0$. Here `lor', an abbreviation of Lorentz, represents the
  4-dimensional differential operator. We study Minkowski's derivation
  and show how these equations generalize to their modern premetric
  form in the framework of tensor and exterior calculus. After
  mentioning some applications of premetric electrodynamics, we turn
  to Minkowski's discovery of the energy-momentum tensor of the
  electromagnetic field. We discuss how he arrived at it and how its
  premetric formulation looks like.\end{abstract}

\maketitle 


\section{Introduction}

In September 1908, Minkowski \cite{CologneAddress} presented a
4-dimensional geometrical version of special relativity theory; for
the historical context, one may compare Walter \cite{ScottWalter}. A
bit earlier \cite{Grundgleichungen}, in December 1907, he gave a
preliminary version, and he expressed Maxwell's theory of
electrodynamics in a 4-dimensional form. For the Maxwell equations he
found, as pointed out by Damour \cite{Thibault} in his historical
analysis, the very compact representation
\begin{equation}\label{MinMax}
\boxed{\text{lor}\, f=-s\qquad\text{and}\qquad \text{lor}\, F^*=0\,.}
\end{equation}
Here lor$_h:=\partial/\partial x_h$ are the components of the
4-dimensional differential operator. The excitation $f$ and the field
strength $F$ are represented in Maxwell's nomenclature
by\footnote{Minkowski took $D\!=\!e$, $H\!=\!m$; $E\!=\!E$, $B\!=\!M$,
  that is, for the excitation $f\sim(e,m)$ and for the field strength
  $F\sim(E,M)$. Even though this seems logical and appropriate ($e,E$
  correspond to electric and $m,M$ to magnetic), the Maxwellian
  notation is so deeply ingrained in our subconsciousness that any
  competing notation seems to have no chance.}
\begin{equation}\label{fF}
( { f}_{hk})= \left(\begin{array}{cccc}
      0\hspace{5pt} &H_z & -H_y & -iD_x\\ -H_z &0\hspace{5pt} & H_x &
      -iD_y\\ H_y & -H_x &0\hspace{5pt} &-iD_z\\ iD_x & iD_y & iD_z
      &0\hspace{5pt}
  \end{array}\right),\quad
(F_{hk})=\left(\begin{array}{cccc}0 &
    B_z & -B_y & -iE_x\\-B_z &0\hspace{5pt} & B_x & -iE_y\\ B_y & -B_x
    &0\hspace{5pt} & -iE_z\\ iE_x & iE_y & iE_z &0\hspace{5pt}
  \end{array}\right),
\end{equation}
with the imaginary unit $i:=\sqrt{-1}$. The rows and columns of the
matrices in (\ref{fF}) are numbered from 1 to 4. The 4-dimensional
electric current is denoted by $s$. The star $^{*}$ used in
(\ref{MinMax})$_2$ is a duality operator, that is,
$F^*_{hk}:=\frac{1}{2}\hat{\epsilon}_{hklm} F_{lm}$, with
$\hat{\epsilon}_{hklm}=\pm 1,0$ as the totally antisymmetric
Levi-Civita symbol and $\hat{\epsilon}_{1234}=+1$; moreover, the
summation convention is applied.  Minkowski demonstrated
\cite{Grundgleichungen} that the equations in (\ref{MinMax}) are
covariant under arbitrary Poincar\'e (inhomogeneous Lorentz)
transformations.

For vacuum (in the ``aether'', as Minkowski said), excitation and field
strength, in suitable units, are related by
\begin{equation}\label{fF0}f=F\qquad\text{or}\qquad
  (f_{hk})=(F_{hk})\,.
\end{equation}
Accordingly, Minkowski found the correct 4-dimensional representation of the
Maxwell equations in the context of special relativity theory.

Later it became clear, perhaps for the first time to Einstein
\cite{Einstein1916}, that the Maxwell-Minkowski equations
(\ref{MinMax}) can be generalized such that they become covariant
under general coordinate transformations (diffeomorphisms) and metric
independent. Then, only the generalization of the constitutive
relation (\ref{fF0}) remains metric dependent.

In the calculus of exterior forms, the 4-dimensional Maxwell equations
for the excitation $H$ and the field strength $\cal F$ eventually
read\footnote{We take the notation of our book \cite{Birkbook}. The
  only exception is that we denote the field strength by $\cal F$ in
  order not to mix it up with Minkowksi's $F$. For other presentations
  of Maxwell's equations in exterior calculus, see, e.g., Bamberg \&
  Sternberg \cite{Bamberg}, Choquet-Bruhat et al.\ \cite{3ladies},
  Delphenich \cite{Dave1,Dave2}, Frankel \cite{Ted}, Jancewicz
  \cite{Bernard}, Lindell \cite{IsmoBook}, Misner, Thorne \& Wheeler
  \cite{MTW}, Russer \cite{Russer}, Thirring \cite{Thirring}, and
  Trautman \cite{Trautman}.}
\begin{equation}\label{MaxExt}
\boxed{dH=J\qquad\text{and}\qquad d{\cal F}=0\,.}
\end{equation}
Here $d$ is the exterior differential. The Minkowskian scheme
(\ref{MinMax}) foreshadows in its compactness the exterior calculus
version in (\ref{MaxExt}). This point was stressed by Damour
\cite{Thibault}. The equations in (\ref{MaxExt}) are
premetric\footnote{We call a structure {\it premetric}, if it can be
  defined ``before'' a metric is available (or a metric exists, but it
  is not used).} and, accordingly, also valid in this form in general
relativity and in general relativistic field theories with torsion and
curvature \cite{Puntigam}. The 2-forms of the excitation $H=\frac
12\,H_{ij}dx^i\wedge dx^j$ and of the field strength ${\cal F}=\frac
12\,{\cal F}_{ij}dx^i\wedge dx^j$, with $i,j=0,1,2,3$, decompose in
1+3 according to \cite{Birkbook}
\begin{equation}\label{HFdecomp}
H=-{\cal H}\wedge dt+{\cal D}\qquad\text{and}\qquad {\cal F}=E\wedge dt +B\,.
\end{equation}
The signs in (\ref{HFdecomp}) are {\it not} conventional. They are
rather dictated by the Lenz rule, see Itin et al.\
\cite{ItinAdP,ItinHehl}. If we put the latter two equations into
matrix form in order to be able to compare them later with the results
of Minkowski, we find
\begin{equation}\label{Gtwiddle}
\hspace{-30pt}H=  (H_{ij})=\left(\begin{array}{cccc}
      0\hspace{5pt} & {\cal H}_1 & {\cal H}_2 & {\cal H}_3\\ -{\cal
        H}_1 &0\hspace{5pt} & {\cal D}^3 &
      -{\cal D}^2\\ -{\cal H}_2 & -{\cal D}^3 &0\hspace{5pt} & 
      {\cal D}^1\\ -{\cal H}_3 & {\cal D}^2 & -{\cal D}^1
      &0\hspace{5pt}
  \end{array}\right),\quad
{\cal F}=({\cal F}_{ij})=\left(\begin{array}{cccc}0 &
    -E_1 & -E_2 & -E_3\\ E_1 &0\hspace{5pt} & B^3 & -B^2\\ E_2 & -B^3
    &0\hspace{5pt} & B^1\\ E_3 & B^2 & -B^1 &0\hspace{5pt}
  \end{array}\right).
\end{equation}
Here ${\cal D}^a:=\frac 12\epsilon^{abc}{\cal D}_{bc}$ and
${B}^a:=\frac 12\epsilon^{abc}{B}_{bc}$, with the 3-dimensional
Levi-Civita symbol $\epsilon^{123}=+1$, that is, $\cal D$ and $B$, as
required by their definitions in physics, are 3-dimensional 2-forms,
whereas $E$ and $\cal H$ are 1-forms. Note that ${\cal D}^a$ and
${B}^a$ are vector {\it densities}.  We stress that the
identifications in (\ref{Gtwiddle}) are premetric, they are valid on
any (well-behaved) differential manifold that can be split locally
into time and space.

The excitation $H$ is a 2-form {\it with} twist (sometimes called odd
2-form) and the field strength $\cal F$ a 2-form without twist
(sometimes called even form 2-form). These are consequences of the
definitions of $H$ and $\cal F$ via charge conservation and the
Lorentz force density, respectively. For this reason, in exterior
calculus, the constitutive law for vacuum (``the spacetime relation'')
reads
\begin{equation}\label{constitext}
H=\lambda_0 \,^\star {\cal F}\,.
\end{equation}
Here $\lambda_0=\sqrt{\varepsilon_0/\mu_0}\approx 1/377\,\Omega$ is
the vacuum admittance; in the units Minkowski chose,
$\lambda_0=1$. The metric-dependent Hodge star operator $^\star$ in
(\ref{constitext}) maps a 2-form without twist into a form with twist, and
vice versa.

In this paper we want to trace the way from Minkowski's system
(\ref{MinMax}) and (\ref{fF}) with (\ref{fF0}) to the premetric
framework (\ref{MaxExt}) and (\ref{Gtwiddle}) with the
metric-dependent spacetime relation (\ref{constitext}). Subsequently,
we analyze Minkowski's discovery of the energy-momentum tensor of the
electromagnetic field and display its modern premetric version.

\section{Minkowski's way to:\; lor $\mathbf{f=-s},$\;
  lor $\mathbf{F^*=0}$}

Minkowski first introduced the fields $f$ and $F$ in Cartesian
coordinates $x,y,z$ and with an imaginary time coordinate $it$ (he put
$c=1$), see (\ref{fF}). Afterwards he changed to the numbering of
coordinates in the following way: $x_1:=x$, $x_2:=y$, $x_3:=z$,
$x_4:=it$. His metric is Euclidean
$ds^2=dx_1^2+dx_2^2+dx_3^2+dx_4^2=g_{hk}dx_h dx_k$, with
$g_{hk}=\text{diag}(1,1,1,1)$. Sometimes he also seems to use the
signature $(-1,-1,-1,-1)$. In any case, as long as Minkowski sticks to
Cartesian coordinates---and this is what he was doing---there is no
need to distinguish contravariant (upper) from covariant (lower)
indices. He displays the conventional Maxwell equations in component
form. The inhomogeneous Maxwell equation reads
\begin{eqnarray}\label{inhom} \nonumber
  \hspace{30pt}\frac{\partial f_{12}}{\partial x_2}
  +\frac{\partial f_{13}}{\partial x_3}
  +\frac{\partial f_{14}}{\partial x_4}&=&s_1\,,\\ \nonumber
  \frac{\partial f_{21}}{\partial x_1}
  +\hspace{30pt}\frac{\partial f_{23}}{\partial x_3}
  +\frac{\partial f_{24}}{\partial x_4}&=&s_2\,,\\ 
  \frac{\partial f_{31}}{\partial x_1}
  +\frac{\partial f_{32}}{\partial x_2}\hspace{30pt}
  +\frac{\partial f_{34}}{\partial x_4}&=&s_3\,,\\ \nonumber
  \frac{\partial f_{41}}{\partial x_1}
  +\frac{\partial f_{42}}{\partial x_2}
  +\frac{\partial f_{43}}{\partial x_3}\hspace{30pt}&=&s_4\,, \nonumber
\end{eqnarray}
and the homogeneous one
\begin{eqnarray}\label{hom} \nonumber
  \hspace{30pt}\frac{\partial F_{34}}{\partial x_2}
  +\frac{\partial F_{42}}{\partial x_3}
  +\frac{\partial F_{23}}{\partial x_4}&=&0\,,\\ \nonumber
  \frac{\partial F_{43}}{\partial x_1}
  +\hspace{30pt}\frac{\partial F_{14}}{\partial x_3}
  +\frac{\partial F_{31}}{\partial x_4}&=&0\,,\\ 
  \frac{\partial F_{24}}{\partial x_1}
  +\frac{\partial F_{41}}{\partial x_2}\hspace{30pt}
  +\frac{\partial F_{12}}{\partial x_4}&=&0\,,\\ \nonumber
  \frac{\partial F_{32}}{\partial x_1}
  +\frac{\partial F_{13}}{\partial x_2}
  +\frac{\partial F_{21}}{\partial x_3}\hspace{30pt}&=&0\,. \nonumber
\end{eqnarray}
Using modern notation, including the summation convention, see Schouten
\cite{Schouten}, we can rewrite (\ref{inhom})
 and (\ref{hom}) as 
\begin{equation}\label{MinMax'}
  \frac{\partial f_{hk}}{\partial x_k}=s_h\qquad\text{and}\qquad
   \frac{\partial F_{hk}}{\partial x_l}
  +\frac{\partial F_{kl}}{\partial x_h}
  +\frac{\partial F_{lh}}{\partial x_k}=0\,.
\end{equation}
Here $h,k,...=1,2,3,4$. Minkowski also introduced the dual of $F_{hk}$
according to $F^*_{hk}:=\frac 12\hat{\epsilon}_{hklm}F_{lm}$, with the
Levi-Civita symbol $\hat{\epsilon}_{hklm}=\pm 1,0$ and
$\hat{\epsilon}_{1234}=+1$. Explicitly, we have
\begin{equation}\label{fF"}
  F^*=(F^*_{hk})=\left(\begin{array}{cccc}0 &-iE_z & iE_y & B_x\\iE_z
      &0\hspace{5pt} & 
      -iE_x & B_y\\ -iE_y & iE_x
      &0\hspace{5pt} & B_z\\ -B_x &-B_y & -B_z &0\hspace{5pt}
  \end{array}\right).
\end{equation}
Then we can put the homogeneous equation in a form reminiscent of the
inhomogeneous one (apart from the sources):
\begin{equation}\label{MinMax''}
\frac{\partial F^{*}_{hk}}{\partial x_k}=0\,.
\end{equation}

Subsequently Minkowski develops a type of Cartesian tensor calculus
with a 4-dimensional differential operator called `lor' (abbreviation
of Lorentz). He introduces ordinary (co)vectors (space-time vectors of
the 1st kind), like $x_h$ and lor$_h:=\frac{\partial}{\partial x_h}$,
and antisymmetric 2nd rank tensors (space-time vectors of the 2nd
kind), like $f_{hk}$ and $F_{hk}$. Then, he can represent
(\ref{MinMax'})$_1$ and (\ref{MinMax''}) symbolically as
\begin{equation}\label{MinMax'''}
\text{lor}\,f=-s\qquad\text{and}\qquad\text{lor} \,F^*=0\,,
\end{equation}
respectively, see (\ref{MinMax}). Using his Cartesian tensor calculus,
Minkowski showed that equations (\ref{MinMax'''}) are covariant
under Poincar\'e transformations.

Einstein and Laub \cite{EL08} thought that Minkowski's presentation of
electrodynamics was very demanding from the point of view of the
mathematics involved. Therefore, they essentially rederived
Minkowski's results in a purely 3-dimensional formalism.

\section{Premetric electrodynamics in tensor calculus}

The next step in the development of the Maxwell equations occurred
immediately after Einstein's fundamental 1915 paper on general
relativity and even before his big survey paper on general relativity
would appear. Now Einstein was in command of tensor calculus in
arbitrary coordinate systems. Einstein \cite{Einstein1916} observed
that Maxwell's equations in vacuum can be put in a generally covariant
form by picking suitable field variables. 

Einstein used the metric $ds^2=g_{\mu\nu}dx^\mu dx^\nu$ with signature
$(1,-1,-1,-1)$, here $\mu,\nu,...=0,1,2,3$. Einstein
\cite{Einstein1916} wrote Maxwell's equations as\footnote{Einstein
  used subscripts for denoting the coordinates $x$, i.e., $x_\tau$
  etc.  Moreover, we dropped twice the summation symbols
  $\Sigma$. Einstein's identifications, which were only worked out for
  vacuum, read \begin{equation}\label{Einstein} {\mathfrak F}=(
    {\mathfrak F}^{\mu\nu})= \left(\begin{array}{cccc} 0\hspace{5pt}
      &E_x & E_y & E_z\\ -E_x &0\hspace{5pt} & H_z & - H_y\\-E_y& -H_z
      &0\hspace{5pt} &H_x\\-E_z & H_y & -H_x &0\hspace{5pt}
  \end{array}\right),\quad
{F}=({F}_{\rho\sigma})=\left(\begin{array}{cccc}0 &
    -E_x & -E_y & -E_z\\E_x &0\hspace{5pt} & H_z & -H_y\\ E_y & -H_z
    &0\hspace{5pt} &H_x\\ E_z & H_y & -H_x &0\hspace{5pt}
  \end{array}\right).
\end{equation} Einstein discussed already earlier the general
covariant form of Maxwell's equations, see
\cite{Einstein1914}. However, the presentation was a bit different 
from (\ref{akademie}) and also by far not as transparent as the one 
in \cite{Einstein1916}. Probably for this reason Einstein 
republished these ideas.}
\begin{equation}\label{akademie}
  \frac{ \partial F_{\rho\sigma}}{\partial x^\tau}+ \frac{ \partial
    F_{\sigma\tau}}{\partial x^\rho}+ \frac{ \partial
    F_{\tau\rho}}{\partial x^\sigma}=0\,,\quad {\frak
    F}^{\mu\nu}=\sqrt{-g}g^{\mu\alpha}
  g^{\nu\beta}F_{\alpha\beta}\,,\quad \frac{\partial{\frak
      F}^{\mu\nu}}{\partial x^\nu}={\cal J}^\mu\,.
\end{equation}
The field strength $F_{\rho\sigma}$ is a tensor, the excitation ${\frak
  F}^{\mu\nu}$ a tensor density. In his ``Meaning of Relativity''
\cite{meaning}, in the part on general relativity, he even picked the
letter $\phi$ for the field strength (instead of $F$) apparently in
order to stress the different nature of $\phi$ and ${\frak F}$.

The Maxwell equations (\ref{akademie})$_1$ and (\ref{akademie})$_3$
are generally covariant {\em and\/} metric independent. Since in
general relativity the metric $g$ is recognized as the gravitational
potential, it is quite fitting that the fundamental field equations of
electromagnetism do {\em not\/} contain the gravitational
potential. The equations in (\ref{akademie}) are specifically valid in
Minkowskian spacetime. Then, of course, the metric occurring in
(\ref{akademie})$_2$ is flat, i.e., its curvature vanishes. Hence by
transforming Minkowski's \,lor $f=-s$ and \,lor $F^*=0$ to curvilinear
coordinates, one can equally well arrive at (\ref{akademie}).

The gravitational potential only enters equation
(\ref{akademie})$_2$. We call this equation the spacetime
relation---it is the ``constitutive law'' of the vacuum. Needless to
say that having understood that there exists a way to formulate
Maxwell's equations in a generally covariant and metric-independent
manner, going back to (\ref{MinMax'}) would appear to be an
anachronism.  Einstein's argument in favor of his formalism, which is
similar to the Minkowski version (\ref{MinMax'}), and not following
(\ref{MinMax'})$_1$ and (\ref{MinMax''}), was that with his formalism
one can derive the energy-momentum tensor of the electromagnetic field
in a much more transparent way. This is certainly true. Still, it was
Minkowski who discovered the energy-momentum expression
\cite{Grundgleichungen} by performing products between the two
space-time vectors of the 2nd kind $f$ and $F$.

But the real decisive step was left to Kottler \cite{Kottler}. He
derived the Maxwell equations from the conservation laws of electric
charge and magnetic flux. The charges in a 3-dimensional volume and the flux
lines piercing a 2-dimensional area are conserved. These two independent
conservation laws are related to counting procedures and thus are
independent of any length or time standard. It is for this reason that
the two Maxwell equations (\ref{akademie})$_1$ and
(\ref{akademie})$_3$ emerge as a priori independent structures and
free of a metric. It is then left to the constitutive law to link the
excitation ${\frak F}^{\mu\nu}$ with the field strength $F_{\a\b}$.

In (\ref{akademie}) we displayed the Maxwellian system in the way
Einstein wrote it. Changing now to modern notation, see Schouten
\cite{Schouten} and Post \cite{Post}, and using the conventions of
\cite{Birkbook}, the premetric Maxwell equations read\footnote{Post
  \cite{Post} denotes $\check{H}^{ij}$ by $\mathfrak{ G}^{\mu\nu}$. In
  Obukhov's investigations on the energy-momentum tensor
  \cite{YuriAnnalen}, the $\check{H}^{ij}$ is simply named ${H}^{ij}$,
  that is, the $\;\check{\null}\;$ is dropped.} ($i,j,...=0,1,2,3$)
\begin{equation}\label{MaxPost}
 \partial_j \check{H}^{ij}=\check{J}^i\,
\qquad\text{and}\qquad \partial_{[i}{\cal F}_{jk]}=0\,.
\end{equation}
The antisymmetrization bracket is defined according to $[ijk] :=\frac
16(ijk-jik+jki-+...)$ i.e., a plus (minus) sign occurs before even
(odd) permutations of the indices $ijk$. Here $\check{H}^{ij}$ are
components of a tensor density, whereas ${\cal F}_{ij}$ are those of a
tensor. For ${\cal F}_{ij}$ we have the identification
(\ref{Gtwiddle})$_2$ and for $\check{H}^{ij}:=\frac
12\epsilon^{ijkl}H_{kl}$, with $H_{kl}$ identified according to
(\ref{Gtwiddle})$_1$. Thus, explicitly,\footnote{If we abandon Minkowski's
  imaginary time coordinate and go over to $x^i$, with $i=0,1,2,3$,
  then excitation and field strength in Minkwowski's framework read
\begin{equation}\label{fF1}
 \tilde { f}=(\tilde { f}_{hk})= \left(\begin{array}{cccc}
      0\hspace{5pt} &D_x & D_y & D_z\\ -D_x &0\hspace{5pt} & H_z &
      - H_y\\-D_y& -H_z &0\hspace{5pt} &H_x\\-D_z & H_y & -H_x
      &0\hspace{5pt}
  \end{array}\right),\quad
\tilde{F}=(\tilde{F}_{hk})=\left(\begin{array}{cccc}0 &
    E_x & E_y & E_z\\-E_x &0\hspace{5pt} & B_z & -B_y\\ -E_y & -B_z
    &0\hspace{5pt} &B_x\\ -E_z & B_y & -B_x &0\hspace{5pt}
  \end{array}\right).
\end{equation}}
\begin{equation}
\label{checkH}
\check{H}=(\check{H}^{ij})= \left(\begin{array}{cccc}
    0\hspace{5pt} & {\cal D}^1 & {\cal D}^2 & {\cal D}^3\\ 
    -{\cal D}^1 &0\hspace{5pt} & {\cal H}_3 &
    -{\cal H}_2\\ -{\cal D}^2 & -{\cal H}_3 &0\hspace{5pt} & 
    {\cal H}_1\\ -{\cal D}^3 & {\cal H}_2 & -{\cal H}_1
    &0\hspace{5pt}
  \end{array}\right).
\end{equation}
If we substitute $\check{H}^{ij}$ into (\ref{MaxPost})$_1$ and use
$J_{ijk}:=\frac 16\hat{\epsilon}_{ijkl}\check{J}^l$, then
(\ref{MaxPost}) can be displayed in the more symmetric form of
\begin{equation}\label{MaxPost'}
 \partial_{[i}H_{jk]}={J}_{ijk}\,
\qquad\text{and}\qquad \partial_{[i}{\cal F}_{jk]}=0\,.
\end{equation}

Today it is not only of academic interest to put Maxwell's equations
in a general covariant form. On the surface of a neutron star, for
example, where we might have strong magnetic fields of some $10^{10}$
tesla and a huge curvature of spacetime---we are, after all, not
too far outside the Schwarzschild radius of the neutron star of some 3
km---the Maxwell equations keep their form (\ref{MaxPost}). The
metric of spacetime enters the vacuum relation, see
(\ref{akademie})$_2$. There is no longer place for only a Poincar\'e
covariant formulation of Maxwell's equations \`a la (\ref{MinMax'}). A
less extreme case is the Global Positioning System. Nevertheless, also
the GPS rules out Poincar\'e covariant electrodynamics, see Ashby
\cite{Ashby}.

\subsection{Constitutive relation}

The system of the $4+4$ Maxwell equations (\ref{MaxPost}) for the
$6+6$ independent components of the electromagnetic field
$\check{H}^{ij}$ and ${\cal F}_{kl}$ is evidently underdetermined. To
complete this system, a constitutive relation of the form
 \begin{equation}\label{Pconst0}
   \check{H}^{ij}= \check{H}^{ij}({\cal F}_{kl})
 \end{equation}
 has to be assumed. The constitutive relation (\ref{Pconst0}) is
 independent of the Maxwell equations and its form can be determined
 by using experimental results. For general {\it magnetoelectric}
 media, including the vacuum as special case, we assume with Tamm
 \cite{Tamm2,Tamm3}, v.~Laue \cite{Laue}, and Post \cite{Post} the
 {\it local}\footnote{A nonlocal generalization of this law is being
   investigated by Mashhoon, see his review \cite{Bahram}.}  and {\it
   linear} premetric relation
\begin{equation}\label{Tamm}
  \check{H}^{ij}=\frac
  12\,\chi^{ijkl}{\cal F}_{kl}\,,
\end{equation}
where $\chi^{ijkl}$ is a {\it constitutive tensor density} of rank 4
and weight $+1$, with the dimension $[\chi]=1/$resistance. Since both
$ \check{H}^{ij}$ and ${\cal F}_{kl}$ are antisymmetric in their
indices, we have $\chi^{ijkl}= -\chi^{jikl}=-\chi^{ijlk}$. An
antisymmetric pair of indices corresponds, in four dimensions, to six
independent components. Thus, the constitutive tensor can be
considered as a $6\times 6$ matrix with 36 independent components.

One can take the premetric Maxwell equations (\ref{MaxPost}) together
with the premetric ansatz (\ref{Tamm}) as an electrodynamic framework
(``local and linear premetric electrodynamics''), see also Matagne
\cite{Matagne}. Therein one can study the propagation of
electromagnetic waves in the geometrical optics approximation. It
turns out, see \cite{Rubilar,Birkbook}, that the premetric, totally
symmetric {\it Tamm-Rubilar tensor}\footnote{Parentheses around
  indices denote total symmetrization, that is,
  $(ijkl):=\frac{1}{24}\{ijkl+jikl+jkil+\cdots\}$. The covariant
  epsilon-system is marked by a hat: $\hat{\epsilon}_{ijkl}=\pm 1,0;\;
  \hat{\epsilon}_{0123}=+1$.}
\begin{equation}\label{G4}   
  {{\cal G}^{ijkl}(\chi):=\frac{1}{4!}\,\hat{\epsilon}_{mnpq}\, 
    \hat{\epsilon}_{rstu}\, {\chi}^{mnr(i}\, {\chi}^{j|ps|k}\, 
    {\chi}^{l)qtu }} 
\end{equation} 
---a new structural element in theoretical physics---controls the
light propagation via a generalized Fresnel equation ${\cal
  G}^{ijkl}q_iq_jq_kq_l=0$. The $q_i$ are the components of the wave
vector of the light.\footnote{This scheme was generalized to Abelian
  gauge and string theory by Schuller, Wohlfahrt, et al., see the
  articles \cite{Schuller,Punzi} and the references given there. Also in
  these generalized schemes a Tamm-Rubilar type tensor plays a
  decisive role.} If birefringence is forbidden, then the light cone
emerges \cite{LammerH}, that is, the metric $g_{ij}$ up to a
(dilation) factor \cite{HOPotsdam}.

Let us come back to $\chi^{ijkl}$. As a $6\times 6$ matrix, it can be
decomposed in its tracefree symmetric part (20 independent
components), its antisymmetric part (15 components), and its trace (1
component). On the level of $\chi^{ijkl}$, this {\it
  decomposition} is reflected in
\begin{eqnarray}\label{chidec}
  \chi^{ijkl}&=&\,^{(1)}\chi^{ijkl}+
  \,^{(2)}\chi^{ijkl}+
  \,^{(3)}\chi^{ijkl}\,.\\ \nonumber 36
  &=&\hspace{11pt} 20\hspace{11pt}\oplus \hspace{11pt}15\hspace{12pt}
  \oplus \hspace{18pt}1\,.
\end{eqnarray}
The third part, the {\it axion} part, is totally antisymmetric and as
such proportional to the Levi-Civita symbol, $ ^{(3)}\chi^{ijkl}:=
\chi^{[ijkl]} ={\a}\, {\epsilon}^{ijkl}$. Note that ${\a}$ is a
pseudoscalar since ${\epsilon}^{ijkl}$ has weight $-1$.  Therefore,
the weight of ${\epsilon}^{ijkl}$ is essential information. The second
part, the {\it skewon} part, is defined according to $
^{(2)}\chi^{ijkl}:=\frac 12(\chi^{ijkl}- \chi^{klij})$.  If the
constitutive equation can be derived from a Lagrangian, which is the
case as long as only reversible processes are considered, then
$^{(2)}\chi^{ijkl}=0$. The {\it principal} part $^{(1)}\chi^{ijkl}$
fulfills the symmetries $ ^{(1)}\chi^{ijkl}= {}^{(1)}\chi^{klij}$ and
$^{(1)}\chi^{[ijkl]}=0$.  The premetric constitutive relation now reads
\begin{equation}\label{constit7}
  { \check{H}^{ij}=\frac
    12\left({}^{(1)}{\chi}^{ijkl}+ {}^{(2)}{\chi}^{ijkl} +{\a}\,
      {\epsilon}^{ijkl}\right){\cal F}_{kl}\,.}
\end{equation}

If we assume the existence of a metric $g_{ij}$, then {\it in vacuum}
the constitutive tensor reduces to
\begin{equation}
\chi{}^{ijkl} = \lambda_0\sqrt{-g}\left(g^{ik}g^{jl} 
- g^{il}g^{jk}\right).\label{chi0}
\end{equation}
This formula is, up to a factor, determined uniquely by the symmetries
of the principal part. On substitution of (\ref{chi0}) into
(\ref{Tamm}), we find the vacuum relation
\begin{equation}\label{Kottler}
  \check{H}^{ij}=\lambda_0\,\sqrt{-g}g^{ik}g^{jl}\,F_{kl}\,,\qquad
  \text{or}\qquad H_{ij}=\frac{\lambda_0}{2}\hat{\epsilon}_{ijmn}
  \sqrt{-g}g^{mk}g^{nl}F_{kl}\,,
\end{equation}
compare (\ref{akademie})$_2$. Note that (\ref{chi0}) belongs to the
principal part $^{(1)}{\chi}^{ijkl}$.

In order to compare (\ref{constit7}) with experiments, we have to
split it into time and space parts. As shown in
\cite{Birkbook,Postconstraint} in detail, we can parameterize the {\it
  principal} part by the 6 permittivities $\varepsilon^{ab}=
\varepsilon^{ba}$, the 6 permeabilities $\mu_{ab}=\mu_{ba}$, and the 8
magnetoelectric pieces $\g^a{}_b$ (the trace vanishes, $\g^c{}_c=0$)
and the {\it skewon} part by the 3 permittivities $n_a$, the 3
permeabilities $m^a$, and the 9 magnetoelectric pieces
$s_a{}^b$. Then, the premetric constitutive relation (\ref{constit7})
can be rewritten as
\begin{eqnarray}\label{explicit'}\nonumber
  {\cal D}^a\!&=\!&\left( \varepsilon^{ab}\hspace{4pt} - \,
    \epsilon^{abc}\,n_c \right)E_b\,+\left(\hspace{9pt} \gamma^a{}_b +
    s_b{}^a - \delta_b^a s_c{}^c\right) {B}^b +
  {\alpha}\,B^a \,, \\ {\cal H}_a\!&=\!&\left( \mu_{ab}^{-1}
    - \hat{\epsilon}_{abc}m^c \right) {B}^b\hspace{2pt} +\left(- \gamma^b{}_a +
    s_a{}^b - \delta_a^b s_c{}^c\right)E_b\hspace{1pt} -
  {\alpha}\,E_a\,.\label{explicit''}
\end{eqnarray}
Here $\epsilon^{abc}= \hat{\epsilon}_{abc}=\pm 1,0$ are the
3-dimensional Levi-Civita symbols, with
$\epsilon^{123}=\hat{\epsilon}_{123}=+1$. As can be seen from our
derivation, ${\alpha}$ is a 4-dimensional pseudo (or axial) scalar,
whereas $s_c{}^c$ is only a 3-dimensional scalar. The cross-term
$\gamma^a{}_b$ is related to the Fresnel-Fizeau effects.  The skewon
contributions $m^c,n_c$ are responsible for electric and magnetic
Faraday effects, respectively, whereas the skewon terms $s_a{}^b$
describe optical activity. Equivalent constitutive relations were
formulated by Serdyukov et al.\ \cite{Serdyukov}, p.86, and studied in
quite some detail.

\subsection{Some applications of premetric structures}

The theory of a possible {\it skewon} part of (\ref{constit7}) lies in
its infancy, see \cite{SkewonPLA,SkewonPRD}.
The {\it axion} part was widely neglected in the literature, if not
even put to zero right away \cite{Post}. As we can see from
(\ref{constit7}) and (\ref{explicit'}), the axion piece obeys
\begin{equation}\label{axion}
  ^{(3)}\!\check{H}^{ij}=\frac 12\a\,\epsilon^{ijkl}{\cal F}_{kl}\,
  \qquad\text{or}\qquad\left\{\begin{array}{l}
      {\cal D}=\hspace{8pt} \a B \\{\cal H}=-\a E  
    \end{array}\right\}.
\end{equation}

\noindent$\bullet$ Tellegen \cite{Tellegen1,Tellegen2} introduced the
{\it gyrator}, with $\a=const$, as a new device in the theory of
two-port networks that ``rotates'' the field strength $(B,E)$ into the
excitation $({\cal D},{\cal H})$---that is, ``voltages'' into
``currents''---and vice versa, see \cite{Postconstraint}.\medskip

\noindent$\bullet$ In the {\it axion-electrodynamics} of Wilczek
\cite{Wilczek}, the axion piece as a field is added to
(\ref{Kottler}),
\begin{equation}\label{Kottler'}
  \check{H}^{ij}=\left(\lambda_0\,\sqrt{-g}g^{ik}g^{jl}
+\frac{1}{2}\a(x)\epsilon^{ijkl}\right)F_{kl}\,,
\end{equation}
see also Itin \cite{YakovAxion}. Moreover, kinetic terms for the axion
are added in order to make the axion a propagating field.\medskip

\noindent$\bullet$ Lindell \& Sihvola \cite{PEMC,PEMCAdP}, again for
$\a=const$, defined the {\it perfect electromagnetic conductor} (PEMC)
via (\ref{axion}). Hopefully the PEMC can be realized by suitable
metamaterials, see Sihvola \cite{metaAri}. Lindell \& Sihvola arrived
at the PEMC when studying electrodynamics in its 4-dimensional
version. Then the minus sign in ${\cal H}=-\a E$ emerges naturally.
Note, ${\cal D}=-sB$ and ${\cal H}=-sE$, with $s:=s_c{}^c$, see
(\ref{explicit'}), represent a (spatially) isotropic skewon that is
only a 3-dimensional scalar. Thus, the relative sign in the second
column of (\ref{axion}) is decisive.\medskip

\noindent$\bullet$ On the basis of experiments in solid state physics,
we were able to show \cite{PRA} that in Cr$_2$O$_3$ a phase can occur
(below the Curie temperature) that obeys the relation (written in
terms of exterior forms) $ ^{(3)}\!H=\a {\cal F}$, with $\a\approx
10^{-4}\lambda_0$. This proves that $H$ is a form with twist and $\cal
F$ one without twist thereby demonstrating the consistency of the
system $dH=J,\; d{\cal F}=0\, \text{and}\, ^{(3)}\!H=\a {\cal F}$.

\section{Premetric electrodynamics in exterior calculus}

One of the principal goals of exterior calculus is to put equations,
in the case under consideration, Maxwell's equations and the
constitutive law, into a manifestly coordinate independent form. In a
way, in exterior calculus, the principle of coordinate covariance is
built-in. If we handle the 2-form $H$, we handle a geometrical quantity
which does not depend on coordinates at all. Only if we are interested
in its components $H_{ij}$, with $H=\frac 12\,H_{ij}dx^i\wedge dx^j$,
then these components are, of course, coordinate dependent. Thus, the
Maxwell equations as such, namely $dH=J$ and $d{\cal F}=0$ are just
relations between forms, coordinates do not play a role. Naturally,
such a calculus expresses the universality of Maxwell's equations in a
particular impressive way.

Let us derive the Maxwell equations in modern language {\it ab ovo} in
order to liberate the physics of Maxwell's equations from historical
ballast, see \cite{Birkbook} for details. Electrodynamics is based on
two conservation laws. First, we have {\it electric charge
  conservation} (first axiom). The charge-current density 3-form $J$
is closed $dJ=0$ and, by de Rham's theorem, also
exact, \begin{equation}\label{Maxinh}dH=J,\end{equation} with the
excitation 2-form $H$ as ``charge-current potential'' \cite{TT}. This
is already the inhomogeneous Maxwell equation.

The current $J$ is a 3-form {\it with} twist ``since an electric
charge has no screw-sense,'' cf.\ Schouten \cite{Schouten}. As a
consequence, the excitation $H$ also carries twist or, in the words of
Perlick \cite{Perlick} (my translation), ``...one {\it must}
understand the excitation as a form with twist if one wants that the
charge contained in a volume always has the same sign, independent of
the orientation chosen.''

We decompose the 4-dimensional excitation $H$ into two pieces: one
along the 1-dimensional time $t$ and another one embedded in
3-dimensional space ($a,b=1,2,3$). We find (see \cite{Birkbook})
\begin{equation}\label{cutexcitation}
  H=-{\cal H}\wedge dt + {\cal D}=-{\cal H}_adx^a\wedge dt+\frac
  12{\cal D}_{ab}dx^a\wedge dx^b
\end{equation}
which, if put in matrix form, yields the identification
(\ref{Gtwiddle})$_1$. Substitution of (\ref{cutexcitation}) into
(\ref{Maxinh}) leads to the 3-dimensional inhomogeneous Maxwell
equations,
\begin{equation}\label{dH}
   \underline{d}\,{\cal D}=\rho\,,\qquad
   \underline{d}\,{\cal H}-\dot{{\cal D}}=j\,,
\end{equation}
i.e., to the Coulomb-Gauss and the Oersted-Amp\`ere-Maxwell laws,
respectively.  The underline denotes the 3-dimensional exterior
differential and the dot differentiation with respect to time.

With charge conservation alone, we arrived at the inhomogeneous
Maxwell equations (\ref{Maxinh}). Now we need some more input for deriving
the homogeneous Maxwell equations. The force on a charge density is
encoded into the axiom of the {\em Lorentz force density\/} (second
axiom)
\begin{equation}\label{Lorentz}
  f_\alpha =(e_\alpha\rfloor {\cal F})\wedge J\,.
\end{equation}
Here $e_\alpha$ is an arbitrary (or anholonomic) frame or tetrad, a
basis of the tangent space, with $\alpha=0,1,2,3$, and $\rfloor$
denotes the interior product (contraction). In tensor language, we can
express (\ref{Lorentz}) as $\check{f}_i={\cal F}_{ik}
\,\check{J}^k$. The axiom (\ref{Lorentz}) should be read as an
operational procedure for defining the electromagnetic field strength
2-form ${\cal F}$ in terms of the force density $f_\alpha$, known
{}from mechanics, and the current density $J$, known {}from charge
conservation. According to (\ref{Lorentz}), the field strength $\cal
F$ turns out to be a 2-form without twist. Its 1+3 decomposition reads
\begin{equation}\label{Fdecomp}
  {\cal F}=E\wedge d\sigma+B=E_adx^a\wedge dt +\frac 12
  B_{ab}dx^a\wedge dx^b\,,
\end{equation}
which yields the identification (\ref{Gtwiddle})$_2$.

{\em Magnetic flux conservation\/} is our third axiom.  The field
strength $\cal F$ is closed, \begin{equation}\label{Maxhom}d{\cal
    F}=0\,.\end{equation} This is the homogeneous Maxwell
equation. Split into 1+3, we find
\begin{equation}\label{dB}
\underline{d}\,E+\dot{B}=0\,,\qquad \underline{d}\,B=0\,,
\end{equation}
that is, Faraday's induction law and the sourcelessness of $B$. The
laws (\ref{dB}) are also premetric. The Lenz rule, as the reason for
the relative sign difference between the time derivatives in
(\ref{dH})$_2$ and (\ref{dB})$_1$, is discussed in
\cite{ItinAdP,ItinHehl}.

We have then the Maxwellian set\footnote{Damour \cite{Thibault}
  discussed Maxwell's equations in exterior calculus in his footnote
  1. He found $\d f=s$ and $\d\star F=0$. The codifferential is defined
  according to $\d:=\star d\star$. For a p-form $\psi$ we have
  ${\star\star}\psi=(-1)^{p-1}\psi$. For the homogeneous equation we
  find $\star d{\star\star}F=-\star dF=0$ or $dF=0$, which coincides
  with (\ref{qaz})$_2$. Thus $F$ can be identified with the field
  strength $\cal F$. We apply the star to Damour's inhomogeneous
  Maxwell equation and find $\star\left(\d
    f\right)={\star\star}d{\star}f=d\star f=S :=\star s$ or $d\star
  f=S$. This equation only agrees with (\ref{qaz})$_1$, provided we
  put $H=\star f$. Consequently, in Damour's exterior calculus, $f$
  loses its operational interpretation as electromagnetic
  excitation. Of course, the codifferential $\d$ is not required for the
  formulation of the Maxwell equations (\ref{qaz}). The exterior
  differential $d$ is sufficient.}
\begin{equation}\label{qaz} dH=J \qquad\text{and}\qquad d{\cal
    F}=0\,.\end{equation} If decomposed into components, we recover
the tensor calculus formulas in (\ref{MaxPost'}). This closes our
considerations on the appropriate form of Maxwell's equations in
4-dimensional spacetime. For vacuum, the Hodge star maps the 2-form
$\cal F$ without twist into the 2-form $H$ with twist according to
\begin{equation}\label{qaz1}
H=\lambda_0\,^\star{\cal F}
\end{equation}
---this is equivalent to (\ref{Kottler}). The Hodge star, as applied
to 2-forms in 4 dimensions, depends only on the conformally invariant
part of the metric, see, e.g., Frankel \cite{Ted} and Kastrup
\cite{Kastrup}. The constitutive relation for axion-electrodynamics
simply reads
\begin{equation}\label{qaz1'}
H=\lambda_0\,^\star{\cal F}+\a{\cal F}\,,
\end{equation}
that is, for the PEMC we have simply $H=\a{\cal F}$. For matter we
have the relation
\begin{equation}\label{qaz2}
H=\kappa({\cal F})
\end{equation}
as analog of (\ref{Pconst0}). If the operator $\kappa$ is local and
linear, we find for its components
\begin{equation}\label{qaz3}
  \kappa_{ij}{}^{kl}=\frac 12\hat{\epsilon}_{ijmn}\chi^{mnkl}\,.
\end{equation}

\section{Minkowski's discovery of the energy-momentum tensor}

Minkowski's ``greatest discovery was that at any point in the
  electromagnetic field {\em in vacuo} there exists a tensor of rank 2
  of outstanding physical importance. \ldots{\em each component of the
    tensor} $E_p{}^q$ {\em has a physical interpretation,} which in
  every case had been discovered many years before Minkowski showed
  that these 16 components constitute a tensor of rank 2. The tensor
  $E_p{}^q$ is called the {\em energy tensor} of the electromagnetic
  field.'' 

  These words of Whittaker \cite{Whittaker}, Vol.\ 2, pp.\ 66 and 67,
  stress the central importance of this discovery. Einstein wouldn't
  have been able to derive the field equations of gravity of general
  relativity without being aware of this important quantity that
  figures as source of his (and Hilbert's) field equation of
  gravity. Einstein's reason for picking the scheme (\ref{akademie}),
  and thus (\ref{MaxPost}) with (\ref{Kottler})$_1$, was that this
  allowed him to derive the energy-momentum tensor of the
  electromagnetic field in a more transparent way than Minkowski.

  Minkowski starts with his ``field vectors'' $f$ and $F$, as
  specified in (\ref{fF}), and multiplies them as matrices according
  to $fF$. The product matrix he decomposes, without giving any
  motivation, into a tracefree piece ${S}$ (that is, with
  $\text{tr}\,{S}=0$), which he calls a spacetime matrix of the 2nd
  kind, and a trace piece ${\cal L}\,\mathbf{1}$, with the unit matrix
  $\mathbf{1}$ and with ${\cal L}:=-\frac 14\text{tr}(fF)$, which
  Minkowski identifies correctly as the Lagrangian density $\frac 12(
  H\!\cdot\! B-D\!\cdot\! E)$ of the electromagnetic field. He wrote,
  if we formulate it compactly in the notation of Cartesian tensor
  calculus,
\begin{equation}\label{matrix}
  S_{hk}=f_{hl}F_{lk}-\frac 14\d_{hk}f_{lm}F_{ml}\,,\qquad\text{that is,}
  \qquad S_{hh}=0\,.
\end{equation}

In order to display the reality relations appropriately, he wrote his
spacetime matrix as
\begin{equation}
  {S}=\left(S_{hk}\right)=\left(\begin{array}{cccc} 
      S_{11}&S_{12}&S_{13}&S_{14}\\
      S_{21}&S_{22}&S_{23}&S_{24}\\
      S_{31}&S_{32}&S_{33}&S_{34}\\S_{41}&S_{42}&S_{43}&S_{44}\end{array}
  \right)=\left(\begin{array}{cccc}X_x&Y_x&Z_x&-iT_x\\
      X_y&Y_y&Z_y&-iT_y\\X_z&Y_z&Z_z&-iT_z\\-iX_t&-iY_t&-iZ_t&T_t
   \end{array} \right).
\end{equation}
With the help of (\ref{matrix}), Minkowski identifies explicitly the
Maxwell stress (momentum flux density)
\begin{eqnarray}\label{MaxStress}
X_x&=&\frac 12\left(D_xE_x-D_yE_y-D_zE_z+H_xB_x-H_yB_y-H_zB_z
 \right)\,,\nonumber\\
X_y&=&D_yE_x+H_xB_y\,,\\
Y_x&=&D_xE_y+H_yB_x\,\quad\text{etc.}\,,\nonumber
\end{eqnarray}
the Poynting vector (energy flux density)
\begin{eqnarray}\label{Poynting}
T_x=H_zE_y-H_yE_z\,\quad\text{etc.}\,,
\end{eqnarray}
and the electromagnetic energy density 
\begin{eqnarray}\label{energy}
  T_t=\frac 12\left(D_xE_x+D_yE_y+D_zE_z+H_xB_x+H_yB_y+H_zB_z \right)\,.
\end{eqnarray}
He did not name the electromagnetic momentum density
$X_t,Y_t,Z_t$---which, by Lebedev \cite{Lebedew}, had already been
discovered experimentally in 1901. Note that Minkowski's
energy-momentum tensor is asymmetric in general and thus, because of
its tracelessness, has 15 independent components (for the algebraic
properties of energy-momentum currents, see Itin
\cite{YakovAlgebraic}).

We leave this historical track and turn first to {\it tensor
  calculus:} Starting with the Lorentz force density
$\check{f}_i={\cal F}_{ij}\check{J}^k$, substituting the inhomogeneous
Maxwell equation (\ref{MaxPost})$_1$, and using (\ref{MaxPost})$_2$,
we arrive at the Minkowski energy-momentum tensor density\footnote{If
  we use $\check{H}^{ij}=\frac 12\epsilon^{ijkl}H_{kl}$, then
  ${\cal T}_i{}^j=\frac{1}{4}\,\epsilon^{jklm}\left(H_{ik}{\cal
      F}_{lm} -{\cal F}_{ik}H_{lm}\right)$.  It is remarkable that
  Ishiwara \cite{Ishiwara} in 1913, inspired by Minkowski's formalism,
  displayed the energy-momentum tensor as $\mathbf{T}=-\frac
  12\left\{\left[\left[\mathbf{H}^*\mathbf{F}^* \right]\right]-
    \left[\left[\mathbf{F}\,\mathbf{H} \right]\right]\right\}$.}
\begin{equation}\label{ff}
  {\cal T}_i{}^{j}=\frac{1}{4}\delta_i^j {\cal F}_{kl}\check{{H}}^{kl}
  -{\cal F}_{ik}\check{{H}}^{jk}\,,\qquad {\cal T}_k{}^k=0\,,
\end{equation}
together with $ \check{f}_i=\partial_j {\cal T}_i{}^{j}+{\cal X}_i$
and the auxiliary force density $ {\cal X}_i= \frac{1}{4}
\left[\left( \partial_i {\cal F}_{jk}\right)\check{{H}}^{jk}- {\cal
    F}_{jk} \,\partial_i\check{{H}}^{jk}\right]$. Incidentally,
already Minkowski proved \cite{Grundgleichungen} that $ {\cal
  T}_i{}^j\,{\cal T}_j{}^k = \frac{1}{4 }\,\delta_i^k\,{\cal T}^2$,
with ${\cal T}^2:={\cal T}_l{}^m{\cal T}_m{}^l$, a formula
that was repeatedly rediscovered.

For {\it vacuum,} $ \check{H}^{ij}=\lambda_0\, \sqrt{-g}g^{ik}g^{jl}\,
F_{kl}$, and thus, in {\it Cartesian coordinates}
[$g_{ij}=\text{diag}(1,-1,-1,-1$)], the auxiliary force density ${\cal
  X}_i=0$. Then ${\cal T}_{ik}:=g_{kj} {\cal T}_i{}^j={\cal T}_{ki}$
is symmetric and conserved, see \cite{Birkbook} for details---and we
recover the result of Einstein \cite{Einstein1916}. At the same time,
under those two conditions (vacuum and Cartesian coordinates),
Minkowski's formula $K=$ lor $S\,$ re-emerges, where $K$ is the
4-dimensional Lorentz force density and $S$ Minkowski's
energy-momentum matrix.

In {\it exterior calculus} we can apply an analogous procedure and
find the energy-momentum 3-form as
\begin{equation}\label{Sigma}
  \Sigma_\alpha :={\frac 1 2}\left[{\cal F}\wedge(e_\alpha\rfloor
    H) - H\wedge (e_\alpha\rfloor {\cal F})\right]\,,\qquad
  \vt^\a\wedge\Sigma_\a=0\,,
\end{equation}
where $e_\a$ denotes the frame (tetrad) and $\vt^\a$ the coframe, with
$e_\a\rfloor\vt^\b=\d_\a^\b$. The expression (\ref{Sigma})$_1$
translates into (\ref{ff})$_1$ via $\Sigma_\a={\cal
  T}_\a{}^\b\,^\star\!\vt_\b$. Accordingly, the formulas (\ref{ff})
and (\ref{Sigma}) incorporate Minkowski's energy-momentum of the
electromagnetic field in a modern notation.

Apparently Minkowski's energy-momentum tensor is defined for matter,
in particular also for {\it moving matter}. Abraham
\cite{Abraham1,Abraham2,Abraham3} introduced a symmetric tensor
instead, since the symmetry was believed to be important from the
point of view of angular momentum conservation, amongst other
reasons. The dispute between the supporters of the Minkowski and the
Abraham tensor is lingering on till today. Obukhov \cite{YuriAnnalen}
reviewed this problem, partly following
\cite{energy-m1,Birkbook,energy-m2}. Obukhov \cite{YuriAnnalen} based
his discussion on a general Lagrange-Noether framework. He recovers
the Minkowski tensor as the {\it canonical} energy-momentum and shows
that the balance equations of energy-momentum and angular momentum are
always satisfied for an open electromagnetic system because of the
asymmetry of the canonical tensor. Moreover, Obukhov is able to
formulate an ``abrahamization'' prescription for symmetrizing a given
asymmetric energy-momentum tensor, provided a timelike vector is
available (a velocity of a medium, for example). Obukhov considers an
ideal fluid with isotropic electric and magnetic properties
interacting with the electromagnetic field. Thereby Obukhov resolves
the perennial Abraham-Minkowski controversy and assigns the
appropriate places for the Minkowski as well as for the Abraham
tensor.

\section{Discussion}

Compare, in vacuum, Minkowski's system
\begin{eqnarray}\label{Minkowski}
  \left\{\text{lor}\right.\! f=\left.\!\!-s,\quad\text{lor}\,F^*=0;
    \quad f=F\right\},
\end{eqnarray}
with the system in exterior calculus
\begin{eqnarray}\label{Exterior}
  \left\{\right.\!\! dH=J,\left.\hspace{-4pt}\qquad d{\cal F}=0; 
\hspace{-4pt}\qquad H={}^\star {\cal F}\right\}.
\end{eqnarray}
The first system is Poincar\'e covariant, the second one generally covariant.

Equation (\ref{Minkowski}) is a special case of
(\ref{Exterior}). Since in Minkowski's formalism \,lor $F^*=0$ in its
version (\ref{MinMax'})$_2$ is already generally covariant, we rewrite
it tentatively as $\,(\text{lor}^*)\,F=0$ and identify
$\text{lor}^*\rightarrow d$. Then, the inhomogeneous equation becomes
$(\text{lor}^*)\,f^*=s\>$ or $\>d\, f^*=s$, and we arrive at
$  \left\{d\,f^*=s,\> dF=0; \>
    f=F\right\}$,
  or at (\ref{Exterior}), provided we identify $H\rightarrow{} f^*$
  and ${\cal F}\rightarrow -F$. Roughly we could say that in the
  transition from (\ref{Minkowski}) to (\ref{Exterior}) the star
  migrated from the Maxwell equations to the constitutive relation for
  the vacuum. Of course, the system (\ref{Minkowski}) is now
  history. The set (\ref{Exterior}), which was derived from a
  reasonable axiomatics of Maxwell's theory, has superseded it. Still,
  we find it remarkable how powerful Minkowski's formalism is and how
  closely (\ref{Exterior}) resembles (\ref{Minkowski}).

And in exterior calculus Minkowski's energy-momentum matrix $S_{hk}$
finds its particular compact expression as energy-momentum 3-form 
\begin{equation}\label{XXX}
  \Sigma_\alpha :={\frac 1 2}\left[{\cal F}\wedge(e_\alpha\rfloor H) -
    H\wedge (e_\alpha\rfloor {\cal F})\right]\,.
\end{equation}
Summing up: The Maxwell equations $dH=J$, $d{\cal F}=0$, the Lorentz
force density $f_\a=(e_\a\rfloor{\cal F})\wedge J$, and Minkowski's
energy momentum `tensor' (\ref{XXX}) represent the framework of
premetric electrodynamics. As such, they are a modern generally
covariant expression of Minkowski's 4-dimensional Poincar\'e covariant
theory of electromagnetism.

\begin{acknowledgement}
  I am very grateful to Yuri Obukhov (Moscow) for thorough and
  detailed discussions on the subject of this article. Moreover, I
  would like to thank the following colleagues and friends for reading
  my article and for coming up with numerous suggestions and
  corrections: Gerhard Bruhn (Darmstadt), Hubert Goenner
  (G\"ottingen), Yakov Itin (Jerusalem), Bernard Jancewicz
  (Wroc{\l}aw), Florian Loebbert (Golm), Bahram Mashhoon (Columbia,
  Missouri), G\"unter Nimtz (K\"oln), Dirk P\"utzfeld (Golm),
  Guillermo Rubilar (Concepci\'on), Tilman Sauer (Pasadena), Gerhard
  Sch\"afer (Jena), Engelbert Sch\"ucking (New York), Ari Sihvola
  (Helsinki), and Yosef Verbin (Tel Aviv). At an earlier stage, this
  work was supported partly by the Deutsche Forschungsgemeinschaft
  (Bonn) with the grant HE 528/21-1.
\end{acknowledgement}

\end{document}